# Structure and Dynamics of Poly(methacrylic acid) and Its Interpolymer Complex Probed by Covalently Bound Rhodamine-123

*Durai Murugan Kandhasamy*[a b*]

[a] National Centre for Ultrafast Processes, University of Madras

Sekkizhar Campus, Taramani, Chennai, INDIA-600113

[b] Present Address: Department of Bioelectronics and Biosensors

Alagappa University, Karaikudi, India-630003

*Email: kdmurugan@gmail.com

ABSTRACT:

The dynamics and structural characteristics of polymethacrylic acid bound rhodamine-123 (PMAA–R123) and its interpolymer complex with poly(vinylpyrrolidone) were investigated using single molecular fluorescence studies. The time resolved fluorescence anisotropy decay of PMAA-R123 at different pH exhibits an associated anisotropy decay behavior characteristic of two different environments experienced by the fluorophore and decays with one shorter and another longer lifetime components. The anisotropy decay retains normal bi-exponential behavior under neutral pH. Fluorescence correlation spectroscopic investigation reveals that the attached fluorophore undergoes hydrolysis under basic condition which results in the release of the fluorophore from the polymer backbone. Shrinkage in the hydrodynamic radius of PMAA is observed on addition of the complementary polymer PVP which is attributed to the formation compact solubilized nanoparticle like aggregates. The size of particle further decreases on the addition of NaCl. The detailed results show that these complexes have potential for use as drug-delivery system under physiological conditions.

KEYWORDS:

Polyelectrolyte, Interpolymer complex, Drug–delivery, Diffusion, fluorescence spectroscopy.

## 1. INTRODUCTION:

Stimuli-responsive polymers, also referred as smart polymers, have attracted considerable research interest due to the capability of dynamic reversible switching of physical and chemical properties upon exposure to external stimuli such as pH, temperature and ionic concentration. Various smart polymers have been synthesized and pursued for practical applications in biomedical imaging, sensing and drug delivery.[1, 2] Among them, polyacrylic acids based polymers, such as poly(acrylic acid), poly(methacrylic acid) and their copolymers, are the most widely studied pH responsive polymers. These water soluble polymers undergoes structural transformation from compact globular form to an extended linear chain at critical pH which shifts towards physiological condition with increasing alkyl chain length at the α-position.[3-5] The ionization of these polymers is a function of pH of the solution and these polymers undergo different conformations regulated by electrostatic interactions, hydrogen bonding, hydrophobic and other weak interactions be connected with pH of the solution.[6-8] The pH sensitive polymer matrices can modulate drug release with respect to acid/base nature of the solution. From the viewpoint of potential applications, pH-sensitive hydrogels seem to be of significance. This is because the physiological pH range is from 1.2 to 7.4 and different body parts operates under specific pH values.[9]

Micelle like polymer particles formed by the self-assembly of interpolymer complexes and amphiphilic block copolymers have been the subject of intensive investigation.[10-11] These studies shows controlled release and improved bioavailability of therapeutic agents which reduced the costs and toxicity of the chemotherapies.[12] The molecular triggers for the drug-delivery system that are responsive to the chemical and physiological signals in the targeting bio-compartments are considered to be the most important and challenging aspect of the drug-delivery system design and synthesis.[13] Earlier, it has been demonstrated that the PMAA based hydrogels absorb and retain polycationic antimicrobial substances and these antimicrobial substances can be released from the PMAA hydrogel in response to changes of pH and salt concentration of the surrounding medium.[14]

PMAA and its copolymers are known to be biocompatible and hold enormous potential for drug delivery system. Moreover, when PMAA and its copolymers are used as a drug delivery agent, they retain hydrophobic, collapsed state in the stomach pH conditions due to the protonation of carboxyl groups. An increase in pH leads to swelling due to carboxyl ionization and hydrogen bond breakage which results the release of loaded drug molecules after gastric passage.[15] In its potential application view point it necessary to understand the dynamics of PMAA and their counter parts in solution of different pH mimicking the physiological environments. The excellent photophysical properties such as high absorption co-efficient, high fluorescence quantum yield, high photostability and relatively long emission wavelength of the dyes based on xanthene scaffolds, mainly rhodamine and fluorescein dyes have attracted considerable interest among the numerous classes of highly fluorescent dyes.[16, 17] The fluorescent dye Rhodamine-123 covalently attached to the polymethacrylic acid is used to probe the conformational transition of polymethacrylic acid in aqueous solution using fluorescence correlation spectroscopy and time-resolved fluorescence anisotropy techniques. In this respect we investigate the structural transition of poly(methacrylic acid) in aqueous solution as function of degree of dissociation of the carboxylic acid group of the polymer back bone. Further, the effect of interpolymer complex formation with PVP on the hydrodynamic radius of PMAA and also the presence of small molecular weight additive, NaCl were investigated using fluorescence correlation spectroscopy.

## 2. EXPERIMENTAL DETAILS:

### Materials and methods:

Analytical grade methacrylic acid and rhodamine-123 were obtained from Sigma–Aldrich chemicals. Potassium peroxydisulphate was obtained from SISCO Research Laboratory. All other reagents were of analytical grade and used as received. The water used for all the experiments was doubly distilled. The methacrylic acid monomer was distilled under vacuum and the middle fractions were collected and stored at 0ºC. Methacrylic acid was polymerized by radical polymerization in aqueous solution. The freshly distilled monomer was polymerized at 60°C under nitrogen atmosphere using 0.1Wt % of potassium peroxydisulphate as radical initiator for about 6 hours. The highly viscous poly(methacrylic Acid) (PMAA) was precipitated in excess of 4:1 acetone petroleum ether mixture as non–solvent, purified by repeated dissolution in water followed by precipitation of excess non–solvent mixture. Thus, obtained PMAA was dried and stored in a vacuum desiccator. The molecular weight of PMAA was determined by gel–permeation chromatography (GPC). Agilent 1200 high–performance liquid chromatography (HPLC) with Agilent ZORBEX GF–250(4.6mmX250mm) column was used for the determination of molecular weight. Polyvinyl alcohol of various molecular weights in phosphate buffer (pH 7.4) was used as reference and making use of UV–Vis Agilent Infinium 1260 Diode Array Detector.

The rhodamine-123 was covalently tagged with PMAA following the procedure reported in the literature (See supporting information, Scheme-S1).[5, 18] Briefly, Carboxyl groups of the polymer and the amino groups of the probe molecules were condensed together to form an amide linkage between the fluorophore and polymer back bone. Required amount of dye in 1% aqueous solution of PMAA was maintained at 90$^{o}$C for 6 hours with stirring. Then the dye molecules which were not covalently bound to the polymer were removed by dialysis of the solution against distilled water for several weeks using cellulose tubing (Sigma–Aldrich) having molecular weight cut–off of 12.5kD. The dialysis was continued till the water outside the membrane showed negligible absorption of R123. Thus obtained dialyzed sample was precipitated and purified by repeated precipitation using excess of non–solvent and finally dried under vacuum. The ratio between the dye (D) and monomer (M) units of the polymer (M/D) was estimated spectrophotometrically as reported in the literature.[18] The molecular weight of the PMAA used in the present study is 86 kDa and monomer to dye ratio of PMAA-rhodamine 123 was estimated to be 4100.

The absorption spectra of the samples were recorded by using the Agilent UV–Visible diode array spectrophotometer. The fluorescence spectra of the samples were recorded using Horiba–Jobin Yvon FluoroMax–4P spectrofluorimeter in a four side polished 10 mm path length quartz cell.

Fluorescence lifetime of the samples are measured using time correlated single photon counting (TCSPC) technique, a digital technique, counts the photons, which are time correlated with the excitation pulse. Picosecond pulsed Light emitting diode (470nm; Horiba–Jobin Yvon) is used as the excitation source. The fluorescence photons from the sample were collected at right angle to the excitation beam. The emitted photons were detected by a MCP-PMT (Hamamatsu R3809U) after passing through the monochromator (f/3) and data collection was carried out by the software (Data Station v2.1) provided by IBH. Repetitive laser pulsing and emitted photon collection produces a histogram of voltage (time) against counts. The counting continued until 10,000 counts were collected in the peak channel. This histogram represents the measured fluorescence decay. For recording the lamp profile, a scatter was placed instead of the sample and the same procedure was repeated. The fluorescence kinetic parameters (lifetime, amplitudes, etc.) are obtained by deconvoluting the excitation and instrument response function from the measured fluorescence decay. The data analysis was carried out by the software provided by IBH (DAS-6) which is based on deconvolution technique using iterative nonlinear least square methods. The quality of fit is normally identified by the reduced $\chi^2$, weighted residual and the autocorrelation function of the residuals.

**Fluorescence Anisotropy Measurements**

Fluorescence anisotropy studies were carried out by measuring the polarized fluorescence decays, $I_{\parallel}(t)$ and $I_{\perp}(t)$. From $I_{\parallel}(t)$ and $I_{\perp}(t)$ decays, the anisotropy decay function r(t) was constructed using the following relation,

$$r(t) = \frac{I_{\parallel}(t) - G I_{\perp}(t)}{I_{\parallel}(t) + 2G I_{\perp}(t)} \longrightarrow \ (1)$$

where $I_{\parallel}(t)$ and $I_{\perp}(t)$ are fluorescence intensity decays for parallel and perpendicular polarizations with respect to vertically polarized excitation light respectively. G is a correction factor for the polarization bias of the detection setup. The G factor was obtained by measuring two additional decays, $I_{HV}(t)$ and $I_{HH}(t)$, and calculating $G = \int I_{HV}(t)dt\ /$ $IHHtdt$.

**Fluorescence correlation spectrometer:**

The autocorrelation curves are recorded using the set-up shown in (See supporting information, Figure-S1). FCS set-up is based on the principle of confocal microscope. The laser beam (514.5nm) of Ar –ion laser is used as the excitation source. A high numerical aperture objective lens (Olympus UPLSAPO 60x/1.20w) focuses the excitation beam into the diffraction limited spot, and effectively collects the fluorescence from the sample. A dichroic mirror separates the fluorescence from the excitation beam and a long pass filter passes appropriate wavelength of fluorescence. The fluorescence through a small pinhole aperture is focused again on a detector. An avalanche photo diode (APD) (Perkin Elmer SPCM-AQRH-13-FC) detector is used as the photon counting detector at visible region because of the high quantum efficiency. Autocorrelation of the fluorescence intensity was obtained by using a FLEX correlator card (FLEX990EM-12D, Correlator.com, USA). Autocorrelation curves were collected in a routine written in LabView. The correlation curves were finally analyzed in Microcal origin software (Origin Pro 8.1). The instrument was calibrated using Rhodamine 6G as standard to measure confocal volume with known diffusion coefficient.

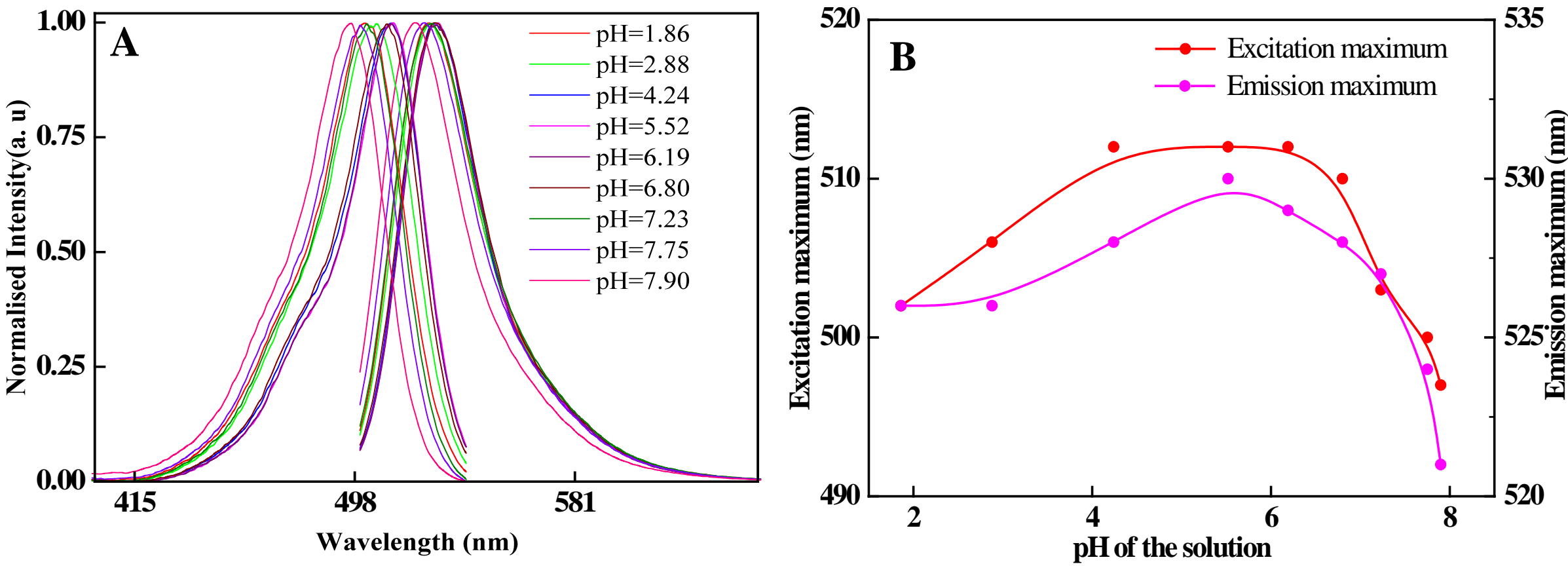

**Figure-1:** A) Excitation ($\lambda_{em}$: 530nm) and emission ($\lambda_{ex}$:490nm) spectra of PMAA bound rhodamine-123 at various pH in aqueous solution. B) Excitation and emission maxima of PMAA bound rhodamine-123 in aqueous solution at various pH.

## 3. RESULTS AND DISCUSSION

### Absorption and emission properties of PMAA-R123:

We have investigated the influence of PMAA environment as a function of pH on the steady state absorption and emission properties of R123. In order to avoid chain entanglement of two different polymer chains, the concentration of solution was maintained below the reported entanglement regime.[19] 20 mM of PMAA-R123 solution was taken to measure the steady state spectral properties of R123 bound to PMAA. Fluorescence excitation spectrum was measured to monitor the absorption spectral changes, and is shown in Fig.1A. The effect of pH on the excitation and emission spectral properties of R123 bound to PMAA is studied at function of pH from 2 to 8 (Figure -1B). The excitation and emission maxima of R123 bound to PMAA at pH 4.0 become red shifted as compared to that of R123 in aqueous solution. It has been observed that both the excitation and emission maxima becomes red shifted with increasing the pH of the solution from 2.0 to 5.0 and further increase in the pH of the solution results in blue shifted excitation and emission maxima. At pH >8 the emission and excitation maximum was found to be similar to that observed in water. It was reported previously that the absorption and emission maxima of rhodamine dyes are shifted to the red region while going from water to alcohols.[20] The present observation is opposite to a normal solvatochromic effect where the excited state is more stabilized in a more polar solvent like water, as evident from characteristic red-shifted absorption and emission bands. This implies that the increasing hydrogen-bonding capability of the amino-group present in the dye molecules with the increase in proton donor ability of protic solvents. The observed spectral shifts are ascribed to the well-known phenomenon called as "blue shift anomaly" which is generally observed in aromatic amines.[21, 22] The protic solvent acts as a hydrogen bond donor to the lone pair of the terminal amino group causing a reduced the involvement of lone pair electrons in the π-conjugation of aromatic moiety. When the dye attached to PMAA, the environment experienced by the dye molecule is polar and/or more exposed to aqueous phase at pH 2.0. On increasing the pH of the solution, the dyes are entrapped into the tightly coiled hydrophobic environment upto pH 5.0. At higher pH, the PMAA chain expanded to more opened rod like structure and the attached dye molecules are exposed entirely to the aqueous environment as revealed from the fluorescence maximum at pH>8.

| pH of the solution | Fluorescence decay parameters | | Anisotropy decay parameters | | | | |
|---|---|---|---|---|---|---|---|
| | $\tau_1$ ns($\alpha_1$) | $\tau_2$ ns ($\alpha_2$) | $r_{0s}$ | $r_{0l}$ | $r_\infty$ | $\phi_s$ (ns) | $\phi_l$ (ns) |
| 1.86 | 0.149(0.05) | 3.98(0.95) | 0.30 | 0.24 | 0.057 | 0.250 | 0.56 |
| 2.88 | 0.147(0.07) | 4.02(0.93) | 0.15 | 0.12 | 0.066 | 0.150 | 6.0 |
| 4.24 | 0.168(0.04) | 4.15(0.96) | 0.22 | 0.17 | 0.080 | 0.250 | 18.5 |
| 5.52 | 0.154(0.06) | 4.11(0.94) | 0.24 | 0.20 | 0.104 | 0.20 | 13 |
| 6.19 | 0.249(0.07) | 4.12(0.93) | 0.24 | 0.21 | 0.105 | 0.150 | 15 |
| 7.23 | --- | 3.95(1.00) | --- | 0.127 | 0.021 | --- | 1.45 |
| 7.75 | --- | 3.86(1.00) | --- | 0.118 | 0.012 | --- | 1.27 |

**Table-1:** Fluorescence decay times and fitted anisotropy parameters of PMAA bound R123 in aqueous solution at various pH. The fluorescence lifetime were obtained by fitting to a bi-exponential decay model. Thus obtained decay times were used to analyze the time resolved anisotropy according to 'associated anisotropy decay model'.

**Fluorescence lifetime studies of PMAA-R123 in aqueous solution:**

The dynamics of radiative relaxation processes of R123 bound to PMAA was characterized by means of time-resolved fluorescence spectroscopy. The free R123 dye in aqueous solution shows a pH independent mono-exponential fluorescence decay curve with the time constant of 4.10 ns. However, the fluorescence decay of R123 attached to PMAA solution with pH less than 7.0 can be fitted with non- exponential functions with at least two different time constants. For this case, the decay curve can be described according to following equation:

$$I(t) = \alpha_1 e^{-(t/\tau_1)} + \alpha_2 e^{-(t/\tau_2)} \longrightarrow \quad (2)$$

where $\alpha_1, \alpha_2$ and $\tau_1, \tau_2$ are the pre-exponential factor and the lifetime of the components, respectively. A relatively good fit by this approach could be characterized by the $\chi^2$ values close to unity or 1±0.2. The lifetimes of the decay components obtained by the deconvolution are listed in Table-1 and the existence of two decay components clearly confirms the presence of two coexistent relaxation pathways and/or emitting species. The longer lifetime component of ~4.0 ns observed for the entire pH range is close to the decay time of the free dye in aqueous solution which reveals that the decay process of these fraction of fluorophores are unaffected by the polymer environment. But the shorter decay component of ~150 ps is due to electron transfer from the polyelectrolyte to the excited fluorophore provided by the close proximity of the dye molecules attached to the carboxylic acid units in aqueous solution due to the globular nature of the polymer under acidic pH conditions. [18, 23, 24]

**Time-resolved fluorescence anisotropy studies:**

Time resolved fluorescence anisotropy measurements were carried out to gain further insight into the dynamics of the PMAA bound R123 as a function of pH. It is generally recognized that the free dyes with single fluorescence decay time in homogeneous medium exhibits single rotational correlation time with fundamental anisotropy value close to 0.4. The exponential anisotropy decay with single rotational correlation time $(\phi)$ can be assumed as a spherical rotor which obeys the following equation [25]

$$r(t) = r_0 e^{-t/\phi} \longrightarrow (3)$$

The dye R123 in aqueous solution shows a single anisotropy decay component with rotational correlation time of 660 picoseconds (Figure-2A). The number of rotational correlation time observed for a molecule that has one fluorescence lifetime depends on the number of independent rotational axes that lead to depolarization. This leads to the proposition that the R123 could be considered as a spherical molecule with single rotational axis. The fundamental anisotropy value $r_0$ for R123 in aqueous solution is around 0.14 and the low value of the fundamental anisotropy indicates the presence of fast rotational diffusion towards the anisotropy decay and is not detected due to the finite time resolution of the TCSPC setup.[26]

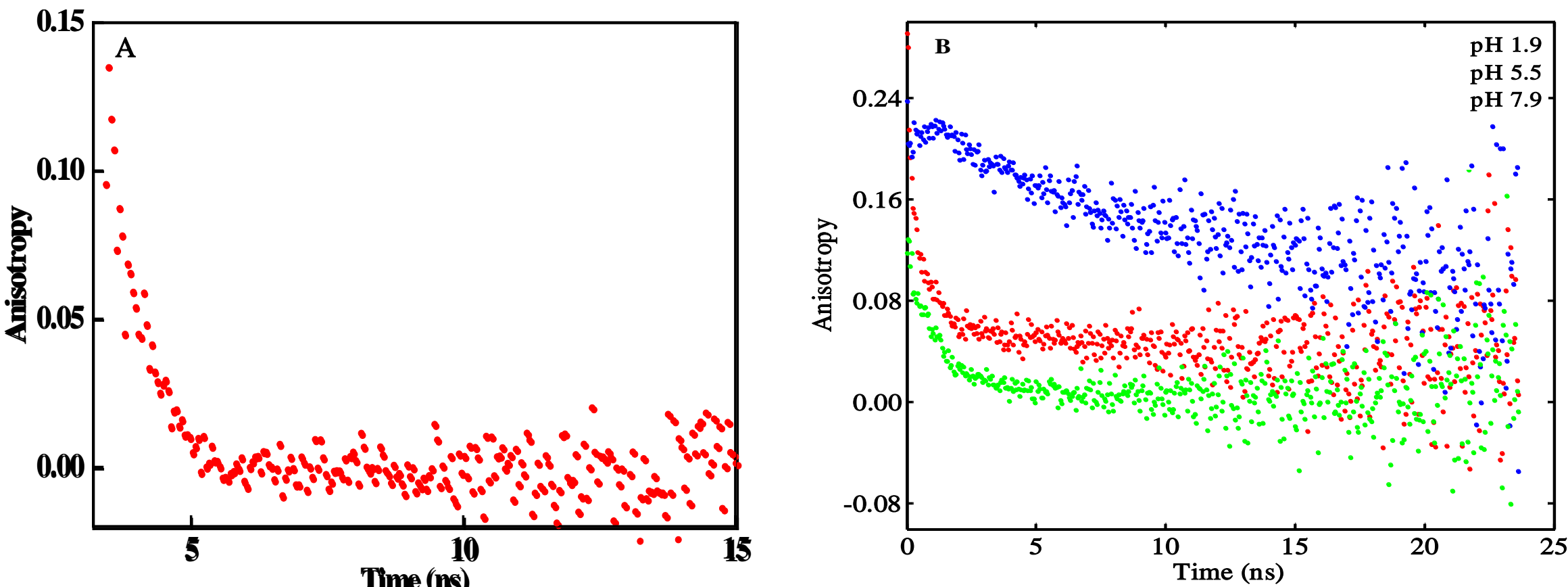

**Figure-2:** Time resolved fluorescence anisotropy decay of A) R-123 in aqueous solution and B) PMAA bound R-123 in aqueous solution at various pH

The anisotropy decay of R123 bound to PMAA exhibits a complex, multi-exponential decay and decay behavior varies with the pH of the solution (Figure-2B). Since the anisotropy decay for heterogeneous systems with multiple fluorescence lifetimes may not be directly related to the rotation of the individual fluorescent species, the analysis of the anisotropy decay is more complex. The anisotropy decay from the mixture of fluorophores is an intensity weighted average of the contribution from the probe in different environment. Such a mixture can display a different type of anisotropy decay called 'associated anisotropy decay' in which the anisotropy decreases to a minimum and then increases at longer times. The analysis can be carried out using an associated heterogeneous model, which assumes that the fluorescent species have different fluorescence lifetimes and different rotational correlation times.[27-29] In this case, an analytical form for the time resolved anisotropy r(t) can be derived from

$$r(t) = f_s(t) r_s(t) + f_l(t) r_l(t) \longrightarrow (4)$$

where $f_s(t)$ is the fraction of fluorescence at time t corresponding to the shorter decaying component, $r_s(t)$ is the anisotropy decay of these component and $f_l(t)$ and $r_l(t)$ are the corresponding quantities for the longer decaying components. It is also evident that the anisotropy decays show the presence of limiting anisotropy factor imparted by the compact nature of the polymer coils and the anisotropy decay can by analyzed using equation-5.

$$r(t) = (r_0 - r_\infty)e^{-t/\phi} + r_\infty \longrightarrow (5)$$

Where, $r_\infty$ is the limiting anisotropy. In the present case, we are dealing with a single rotational correlation time and a single fluorescence decay time for each of the two decaying components of the probe, the combination of fractional terms of fluorescence lifetime and anisotropy terms yields (Equation-6)

$$r(t) = \frac{\alpha_s e^{(-t/s)}((r_{0s} - r_\infty)e^{(-t/\phi_s)} + r_\infty) + \alpha_l e^{(-t/\tau_l)}((r_{0l} - r_\infty)e^{(-t/\phi_l)} - r_\infty)}{\alpha_s e^{(-t/\tau_s)} + \alpha_l e^{(-t/\tau_l)}} \longrightarrow (6)$$

The observed anisotropy decays with complex behavior up to the pH >7.0 are analyzed using the above equation and obtained the values from the best fit are given in Table-1. The fundamental anisotropies ($r_{0s}$ & $r_{0l}$) and the rotational correlation times ($\phi_s$ & $\phi_l$) of the individually decaying components as a function of pH of the solution are presented. The two rotational correlation times were interpreted to be associated with the local motion of R123 (shorter correlation time, $\phi_s$) and the segmental motion of polymer (longer correlation time, $\phi_l$).[30] The longer rotational correlation time changes with pH of the solution which indicates the opening of polymer coils to linear chain. This species with shorter rotational correlation time undergoes a significant decrease in its initial anisotropy upon increasing the pH of the solution, indicating an enhanced mode of fast depolarization due to opening of polymer segments. The two rotational correlation time extracted from the analysis may be attributed to the two different environment of R123 bound to PMAA. The rotational correlation time of the shorter component more or less remains the same and it disappears when the pH of the solution exceeds 6.0. The rotational correlation time for the longer component increases from 6.0 ns to 18.5 ns with increasing pH of the solution till 4.20. Then it starts decreasing gradually and reaches 1.2 ns when the pH of the solution is more than 7.0. It is established that the PMAA forms a more compact structure at pH ~4.5 in aqueous solution due to the pronounced hydrogen bonding between un-dissociated and dissociated carboxylic acid groups in addition to the hydrophobic interaction provided by the $\alpha$-methyl substituent.[18] The observed 3-fold increase in the rotational correlation time upon increasing the pH of the solution to 4.5 disclose the formation more rigid globular structure at this pH. At higher pH single rotational correlation time is observed corresponding to more exposed probe molecule as a result of expansion of the polymer chain to a rod like structure. When, we revert the pH of the solution from 8.0 to 4.5, the longer rotational correlation time was found to be 5.2 ns. This indicates that the dye molecules are hydrolyzed under basic conditions and depart from the polymer matrices when they are in the extended form. The reversible switching the polymer to the globular structure provides a less polar environment to the dye molecules in their close vicinity. The release of covalently attached dye molecules was further confirmed by using a solution of dye molecules polystyrene sulphonate (PSS) in water at different pH (See supporting information, Figure-S2). The longer rotational correlation time was estimated to be close to 5.1 at pH 2.0 and 11.0. The negative sulphonate groups non-covalently attracted the positively charged dye molecules and restrict the motion of the dye molecule. However, the motion is highly restricted when the dye molecules covalently attached to PMAA compared to non-covalently interacting PSS chains. Moreover, the globular structure of PMAA under acidic pH affords a more compact environment to the dye molecules buried inside polymer coils. This is manifested by the slower decay of the longer rotational correlation time of R123 attached PMAA compared to dye dissolved in PSS solution of identical conditions.

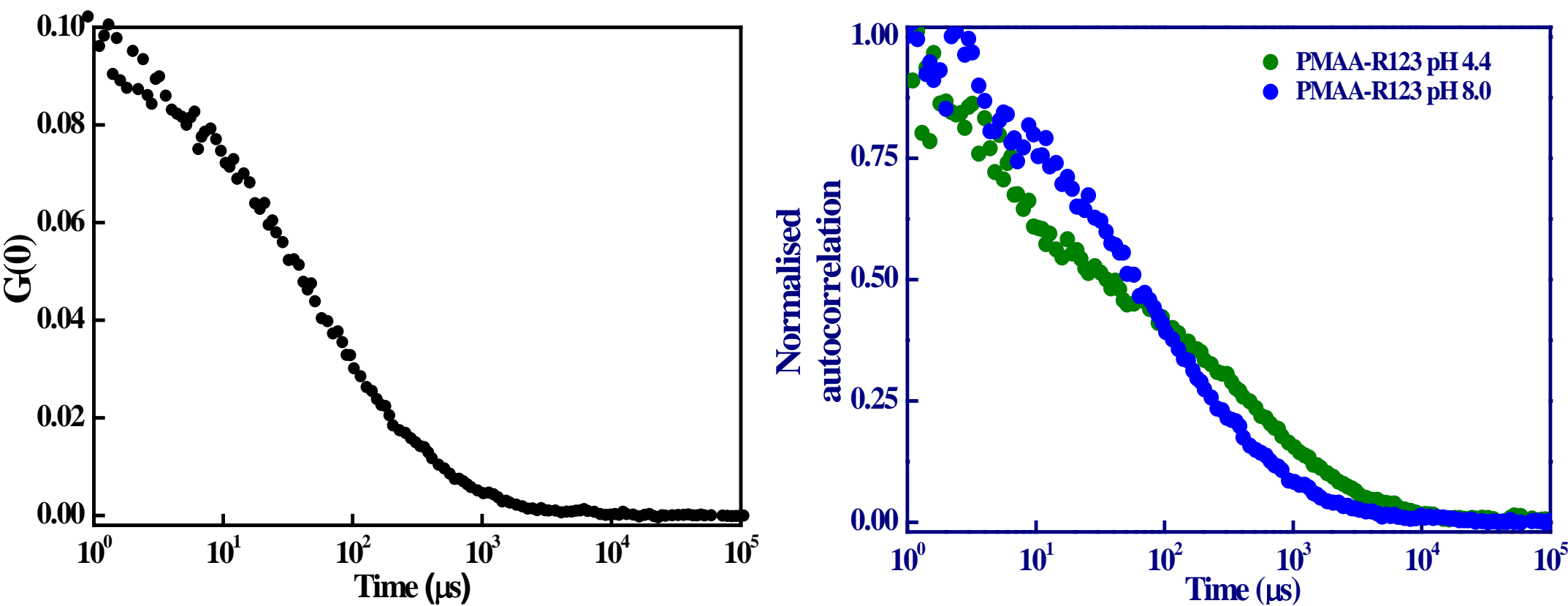

**Figure-3:** A) Autocorrelation curves of rhodamine-123 in aqueous solution and B) PMAA bound rhodamine-123 in aqueous solution at pH 4.4 (green colour) and 8.0 (blue colour)

**Fluorescence correlation spectroscopic studies of PMAA-R123 in aqueous solution:**

The autocorrelation curves of R123 in aqueous solution recorded by using 514 nm Ar-ion laser (5 mW) and are shown in Figure-3A. The estimated diffusion time for R123 is found to be 60 μs. The autocorrelation curves are recorded for R123 bound to PMAA in aqueous solution at various pH and the representative autocorrelation curves at pH 4.40 and 8.0 are shown in Figure-3B.

| pH | Diffusion time (μs) | Fraction of triplet contribution | Triplet decay time(μs) | Diffusion coefficient ($\mu m^2/s$) | Hydrodynamic Radius (nm) |
|---|---|---|---|---|---|
| 2.90 | 216.49 | 0.412 | 5.90 | 91.2 | 2.69 |
| 3.75 | 281.15 | 0.457 | 6.16 | 70.2 | 3.49 |
| 4.40 | 388.54 | 0.475 | 6.64 | 50.8 | 4.82 |
| 5.15 | 322.11 | 0.494 | 5.43 | 61.3 | 4.00 |
| 6.45 | 116.24 | 0.304 | 3.68 | 169 | 1.44 |
| 8.00 | 115.03 | 0.268 | 4.82 | 171 | 1.43 |
| 9.00 | 71.88 | 0.283 | 4.00 | 274 | 0.89 |
| 9.85 | 73.00 | 0.200 | 4.20 | 270 | 0.90 |
| 10.4 | 75.08 | 0.222 | 3.82 | 262 | 0.93 |

**Table-2:** The extracted diffusion time, diffusion coefficient and hydrodynamic radius of rhodamine-123 bound to PMAA in aqueous solution at various pH. The triplet decay time and its contribution in the decay of autocorrelation curve are also presented.

Since the autocorrelation curves shows significant triplet fraction, the auto correlation curves are analyzed based on 3D diffusion with reversible reaction model. From the diffusion times, the diffusion co-efficient and hydrodynamic radius of R123 bound to the polymer are estimated and presented in Table-2. The diffusion time of PMAA-R123 is found to be ~400 μs at pH 4.4 which is six times higher than that of free R123 in aqueous solution. The diffusion time of PMAA-R123 is dependent on pH of the solution. The diffusion time increases with increasing the pH of the solution from 2.0 to 5.0 and further increase in pH results a sudden fall in the diffusion time. The increase in the diffusion time of R123 bound to PMAA indicates the slow diffusion of polymer coils. Pristinski *et.al* studied the conformational transition of PMAA labeled with Alxa-488 using fluorescence correlation spectroscopy in aqueous solution and reported that the diffusion time and hydrodynamic radius increases with the increase in the pH of the solution.[19] The studies were also repeated using phosphate buffer and the results shows identical behavior as a function of pH of the solution. From these results, it is confirmed that the decrease in the diffusion time upon increasing the pH of the solution is due to the release of the attached dye molecule to the aqueous solution.

The dynamic change in the hydrodynamic radius of PMAA upon interpolymer complex formation with PVP has also been investigated with increasing molar ratio of PVP (Figure-4A&B). A decrease in the hydrodynamic radius was observed upon the addition of PVP until the molar ratio of unity. The hydrodynamic radius remains the same on further addition of PVP which indicates the 1:1 complex formation between PMAA and PVP.[31] The decrease in the hydrodynamic radius is due the formation of more compact globules on interpolymer complexation.[18] Further decrease in the hydrodynamic radius observed upon addition of NaCl with 1:1 PMAA/PVP interpolymer complex (Figure-4C). Small molecule additives such as NaCl are known to affect the interpolymer complex formation owing to its salting out nature and results in a more compact rigid particles.

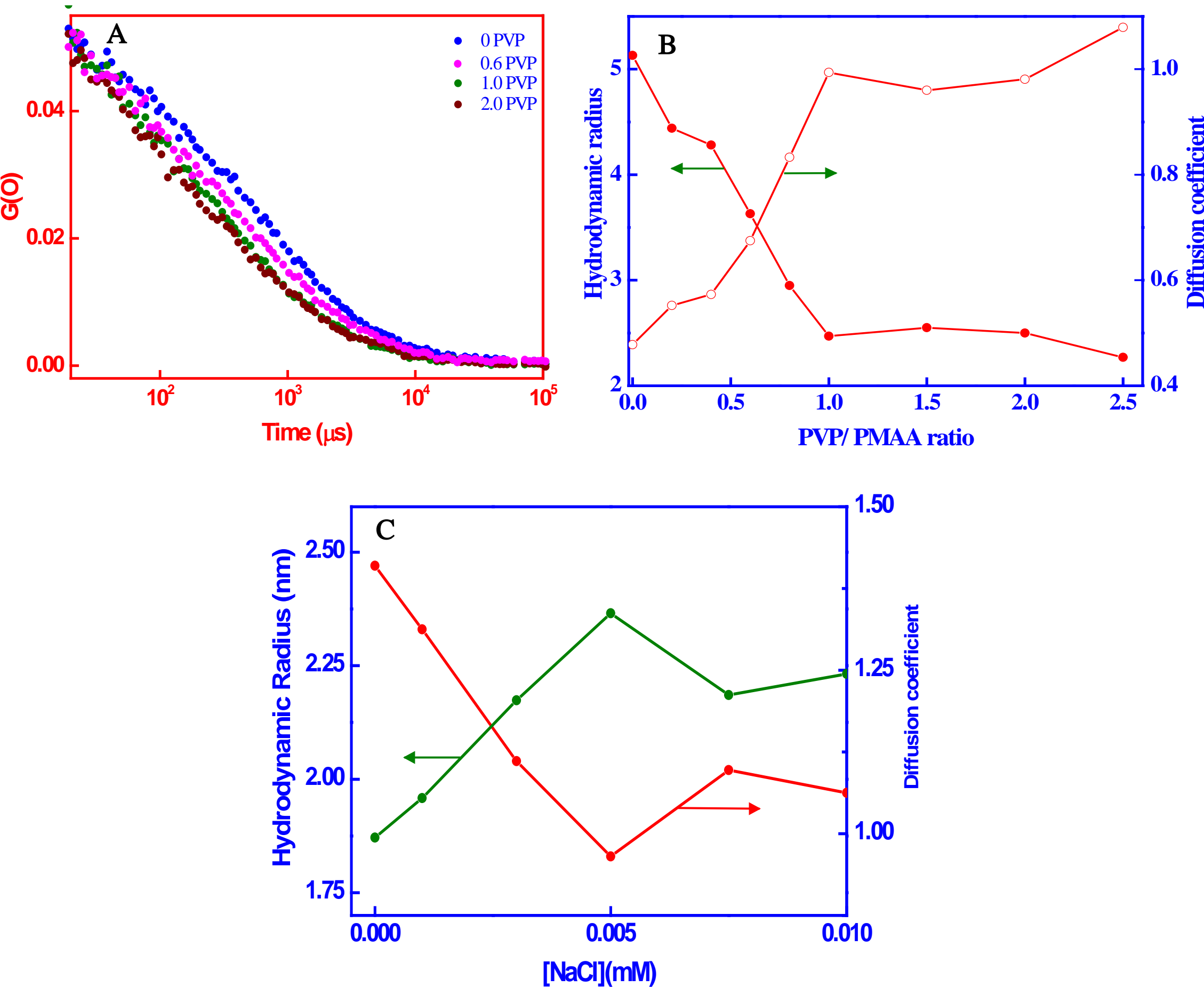

**Figure-4:** A) Correlation curves of Rhodamine-123 attached PMAA with increasing molar ratio of PVP and B) the characteristics of diffusion coefficient, hydrodynamic radius with increasing PVP molar ratio. C) The diffusion coefficient and hydrodynamic radius in the presence of NaCl for the 1:1 interpolymer complex of PMAA/PVP.

The present study may have potential in drug delivery by releasing the attached drug molecules using pH as the stimulating agent. Earlier, we observed the formation of nanoparticles through aggregation of the interpolymer complexes of PMAA/PVP when they dried under ambient conditions.[32] Also, a reduction in particle size was observed upon the addition of NaCl (See supporting information, Figure-S3). Based on the observations reported in this article and our earlier studies a model drug-delivery system can be designed as shown in Scheme-1. The advantages of the present model are the polymers used are biocompatible and water soluble. Moreover, PMAA and its interpolymer complexes can retain the loaded small molecules in physiological conditions when the pH is below 6.0. It can be formulated in a powdered form when it is properly tried. This will be further investigated using different kind of fluorescent molecules, nature and mode of attachment.

**Scheme-1:** Various structural aspects of the model drug delivery system

## 4. CONCLUSION:

The conformational dynamics of fluorescently labeled PMAA is investigated using fluorescence correlation spectroscopy and time resolved anisotropy studies. The present studies reveal that at lower pH the polymer exists as compact coils connected by extended part of the linear chain The anisotropy studies at higher pH shows the detachment of covalently attached fluorophore further supported by the decrease in diffusion time. Interpolymer complex formation made the globular polymer structure more compact and assembled as nanoparticles under certain conditions which will be useful in designing novel polymer based drug delivery system using pH as the stimulating agent.

**Acknowledgement:**

The author thanks to P. Natarajan, former INSA Senior Scientist for his help in designing experiment and his extensive input to improve the content of the manuscript. The author K D acknowledges the financial assistance received from CSIR as Senior Research fellowship and Alagappa University for the postdoctoral fellowship through RUSA-2.

Table of Content:

Structure and Dynamics of Poly(methacrylic acid) and Its Interpolymer Complex Probed by Covalently Bound Rhodamine-123

*Durai Murugan Kandhasamy* [a,b*]

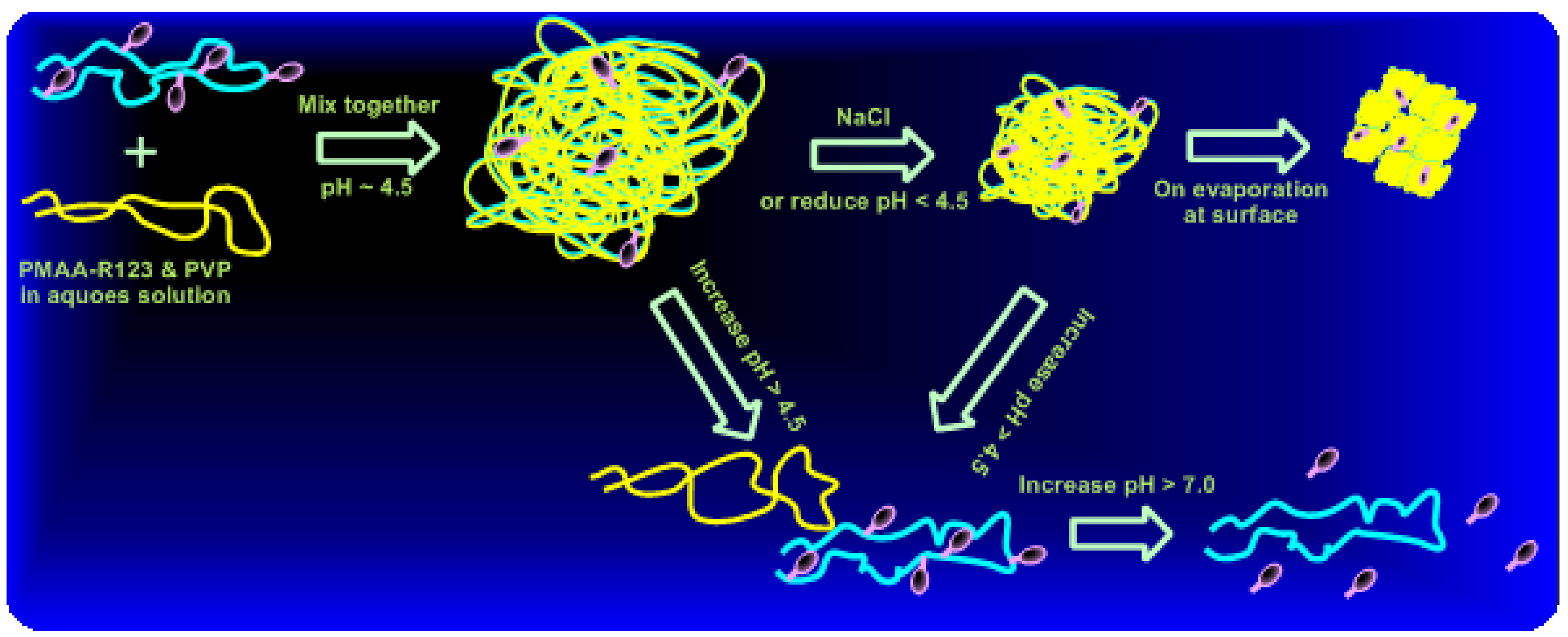